\documentclass[a4paper]{article}

\usepackage{INTERSPEECH2019}

\usepackage{CJKutf8}

\title{Disambiguation of Chinese Polyphones in an End-to-End Framework with Semantic Features Extracted by Pre-trained BERT}
\name{Dongyang Dai$^{1, 2}$, Zhiyong Wu$^{1,2,3,*}$\thanks{* Corresponding author}, Shiyin Kang$^{4}$, Xixin Wu$^3$, Jia Jia$^{1,2}$, Dan Su$^{4}$, Dong Yu$^{4}$,\\Helen Meng$^{1,3}$}
\address{
  $^1$Tsinghua-CUHK Joint Research Center for Media Sciences, Technologies and Systems, \\
  Graduate School at Shenzhen, Tsinghua University, Shenzhen, China\\
  $^2$Beijing National Research Centre for Information Science and Technology (BNRist),\\
  Department of Computer Science and Technology, Tsinghua University, Beijing, China \\
  $^3$Department of Systems Engineering and Engineering Management, \\
  The Chinese University of Hong Kong, Shatin, N.T., Hong Kong SAR, China\\
  $^4$Tencent AI Lab, Tencent, Shenzhen, China}
\email{ddy17@mails.tsinghua.edu.cn, \{zywu,wuxx,hmmeng\}@se.cuhk.edu.hk\\
	\{shiyinkang,dansu,dyu\}@tencent.com, jjia@tsinghua.edu.cn}

\begin{document}

\maketitle
\begin{abstract}
Grapheme-to-phoneme (G2P) conversion serves as an essential component in Chinese Mandarin text-to-speech (TTS) system, where polyphone disambiguation is the core issue. In this paper, we propose an end-to-end framework to predict the pronunciation of a polyphonic character, which accepts sentence containing polyphonic character as input in the form of Chinese character sequence without the necessity of any preprocessing. The proposed method consists of a pre-trained bidirectional encoder representations from Transformers (BERT) model and a neural network (NN) based classifier. The pre-trained BERT model extracts semantic features from a raw Chinese character sequence and the NN based classifier predicts the polyphonic character's pronunciation according to BERT output. In out experiments, we implemented three classifiers, a fully-connected network based classifier, a long short-term memory (LSTM) network based classifier and a Transformer block based classifier. The experimental results compared with the baseline approach based on LSTM demonstrate that, the pre-trained model extracts effective semantic features, which greatly enhances the performance of polyphone disambiguation. In addition, we also explored the impact of contextual information on polyphone disambiguation.
\end{abstract}


\section{Introduction}

Text-to-speech (TTS) technology has been widely used in voice-assistants, car navigation, e-books and other products. For language based on graphic symbols like Chinese, it is necessary to convert the input character sequence into phoneme sequence before synthesizing speech. Therefore the grapheme-to-phoneme (G2P) conversion component is essential in Mandarin TTS system.

\begin{figure}[t]
	
	\begin{minipage}[b]{1.0\linewidth}
		\centerline{\includegraphics[width=8cm]{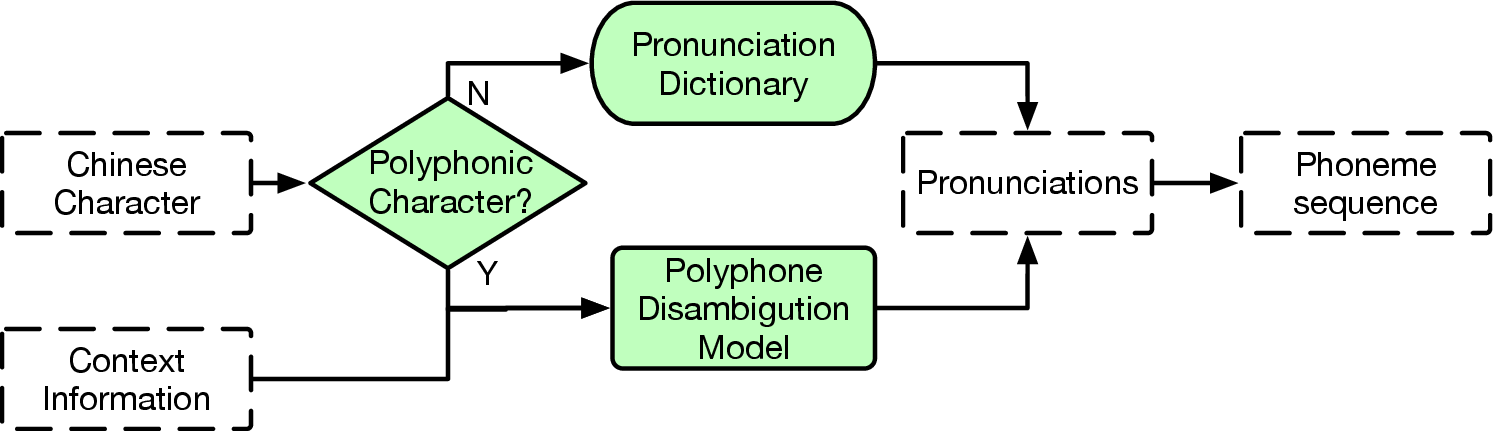}}
		%  \vspace{2.0cm}
	\end{minipage}
	\caption{Chinese G2P conversion flow}
	\label{fig:chineseG2P}
\end{figure}

A Chinese character may have multiple corresponding pronunciations, which is called polyphonic character. Polyphone disambiguation which predicts the correct pronunciation of a polyphonic character is the core issue in Chinese G2P conversion. Fig.\ref{fig:chineseG2P} depicts the flow of Chinese G2P conversion. If the input character is not a polyphonic character, we can directly look up the dictionary to derive its pronunciation. Otherwise, we need a polyphone disambiguation model to predict its pronunciation based on its context information.

For Chinese polyphonic characters, their pronunciations are affected by the semantic context information \cite{lilinhui2010} of neighboring characters that may occur before or after the polyphonic character with different spans. The earliest Chinese polyphone disambiguation system was based on manual rules \cite{dajunzhang2000, lianhongcai1995}. The laws for polyphone disambiguation were summarized by linguistic experts  and written into computer-understandable forms. However, as the number of rules increases, a polyphonic character may be matched by multiple conflicting rules at the same time.

As the amount of data increases, more and more researchers tried to use statistical approaches for polyphone disambiguation. Decision trees were applied in \cite{wang1996broad} to classify the pronunciations of polyphonic characters. A study in \cite{fangzhouliu2007} has shown that a Maxent model outperforms decision tree. However, these statistical approaches need handcrafted features extracted from sentence containing the polyphonic character as model's input. Feature engineering requires linguistic background knowledge and is expensive.

Deep neural network (DNN), which can learn high-level invariant features from raw data \cite{bengio2013representation}, provides an easier way to extract valid features and predict the pronunciation of polyphonic character. Shan et al. \cite{shan2016bi} addressed the polyphone disambiguation problem as a sequential labeling task, and proposed a bidirectional long short-term memory (BLSTM) approach to predict the pronunciation of polyphonic character which outperforms the max entropy model. This approach encodes the polyphonic character’s surrounding observations including neighbor characters and part-of-speech (POS) of neighbor words. Only word tokenization and POS tagging are required in the preprocessing stage in Shan's approach, greatly reducing the work of feature engineering. However, since this model is trained on limited annotation data, it is difficult to learn enough semantic information for polyphone disambiguation. Besides, all the considered polyphonic characters share the same classifier with only one output layer listing the labels of all possible pronunciations, which cannot avoid being predicted to the pronunciation of another polyphonic character, nor can it handle new polyphonic characters that have not yet appeared without changing the output layer and retraining of the shared classifier.

With the development of neural network research, end-to-end TTS has become a new trend \cite{sotelo2017char2wav, wang2017tacotron, li2018close}. Towards end-to-end G2P conversion in Chinese Madarin, we propose an end-to-end framework for polyphone disambiguation consisting of a pertained BERT \cite{devlin2018bert} and neural network (NN) based classifier. The advantage of proposed method is summarized as follows:
\begin{enumerate}
	\item The proposed method predicts pronunciation in an end-to-end way, accepting raw Chinese character sequence containing polyphonic character as input, without the necessity of any preprocessing procedures.
	\item A large amount unsupervised data can be adopted to pre-train the model for extracting semantic information, which will boost the performance of polyphone disambiguation.
	\item The proposed method uses a non-shared output layer among different polyphonic characters, eliminating the case of mis-predicting to pronunciation of other polyphonic characters. Furthermore, with this architecture, when new polyphonic characters are required to be processed, only output layers for these new characters are added and trained without affecting the existing classifiers.
\end{enumerate}

\section{The proposed approach}

\begin{figure}[t]
	\centering
	\includegraphics[scale=0.45]{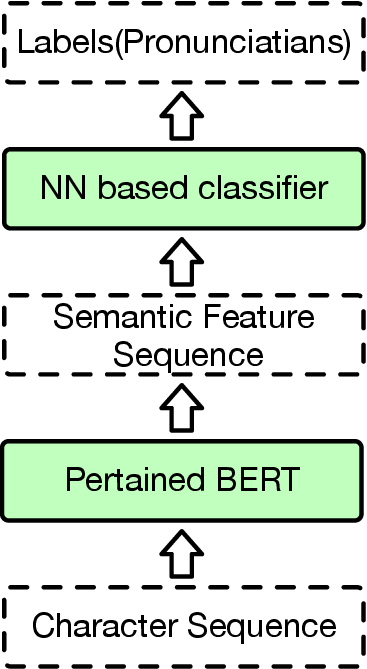}
	\caption{Model architecture}
	\label{fig:model_architecture}
\end{figure}

The proposed framework consists of a pre-trained BERT and NN based classifier. Depicted in  Fig.\ref{fig:model_architecture}, the pertained BERT extracts semantic features from a raw Chinese character sequence containing polyphonic character, the following NN based classifier predicts polyphonic character's pronunciation according to BERT output. In our research, we explored the performance of classifiers based on fully-connected network, BLSTM and Transformer block respectively.

\subsection{The pre-trained BERT}

\begin{figure}[t]
	\centering
	\includegraphics[scale=0.35]{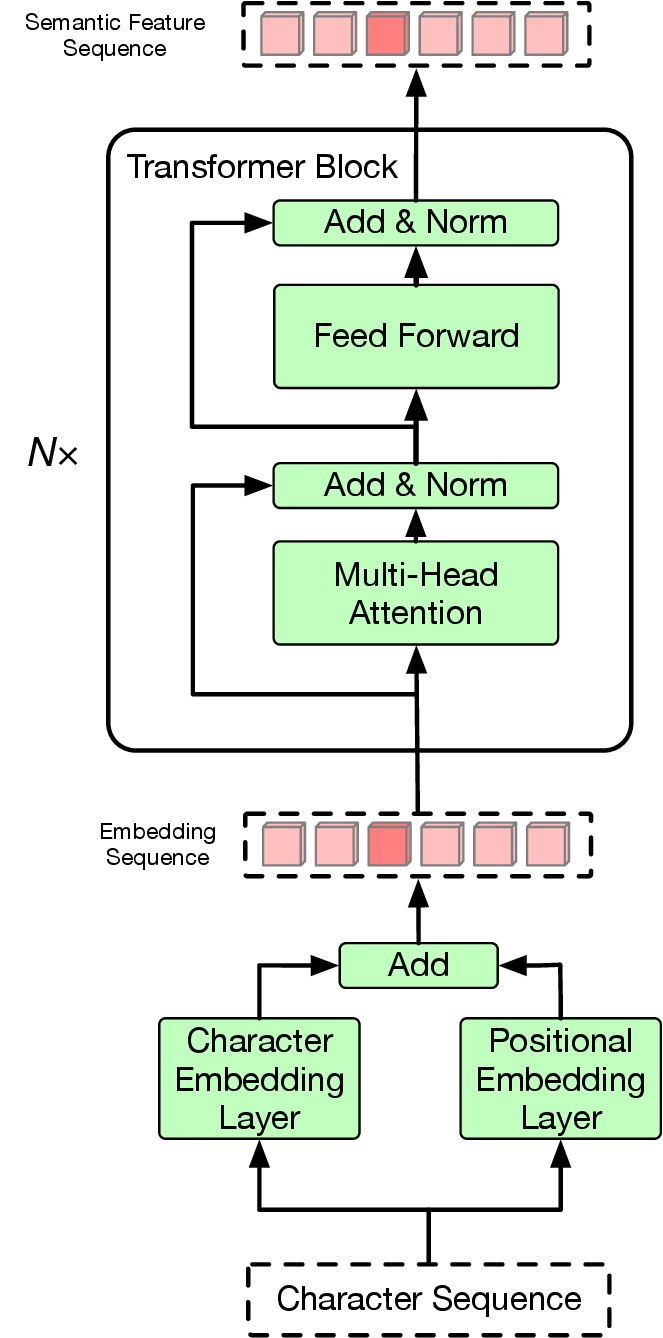}
	\caption{BERT architecture}
	\label{fig:bert}
\end{figure}

The pre-trained BERT accepts raw Chinese character sequence as input and outputs a sequence of semantic features. The BERT's architecture is shown in Fig.\ref{fig:bert}, a character embedding layer and positional embedding layer process the input character sequence respectively before getting the embedding sequence. The following Transformer blocks convert embedding sequence to semantic feature sequence. Because the use of Transformer \cite{vaswani2017attention} and BERT have been ubiquitous, the structure of Transformer block will not be described in detail here.

The BERT model is pre-trained on a large amount of unlabeled data with two prediction task, predicting the masked input characters and predicting the next sentence. The pre-trained BERT model is expected to learn semantic representings from raw character sequence.

\subsection{The NN based classifier}
\begin{figure*}[htbp]

	\begin{minipage}[t]{0.3\linewidth}
		\centering
		\centerline{\includegraphics[width=1.6in]{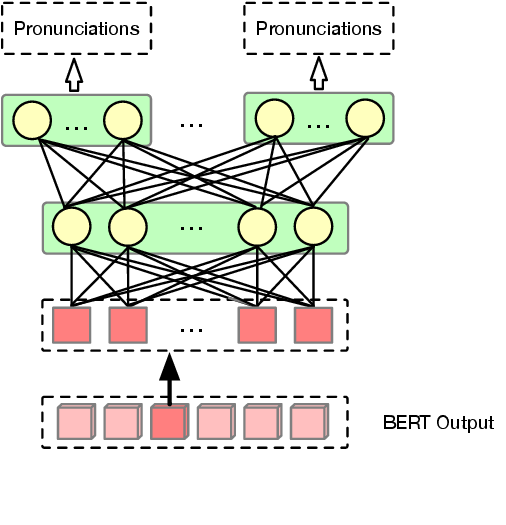}}
		\centerline{(a) \quad\quad}\medskip
		
	\end{minipage}%
	\begin{minipage}[t]{0.3\linewidth}
		\centering
		\includegraphics[width=1.8in, height=2.0in]{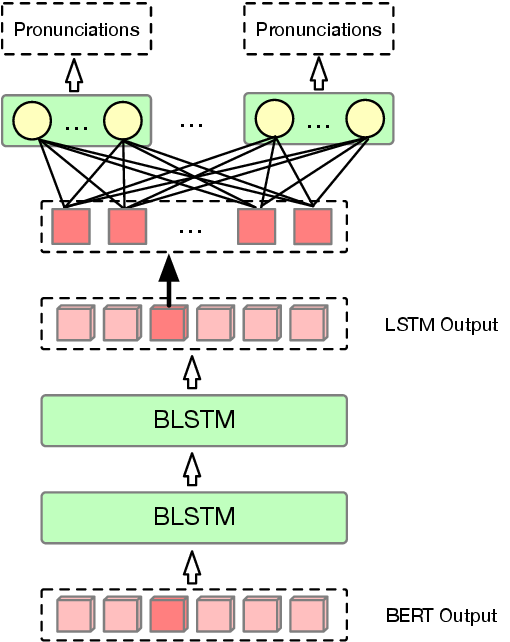}
		\centerline{(b)\quad\quad}\medskip
		
	\end{minipage}
	\begin{minipage}[t]{0.3\linewidth}
		\centering
		\includegraphics[width=1.8in, height=2.0in]{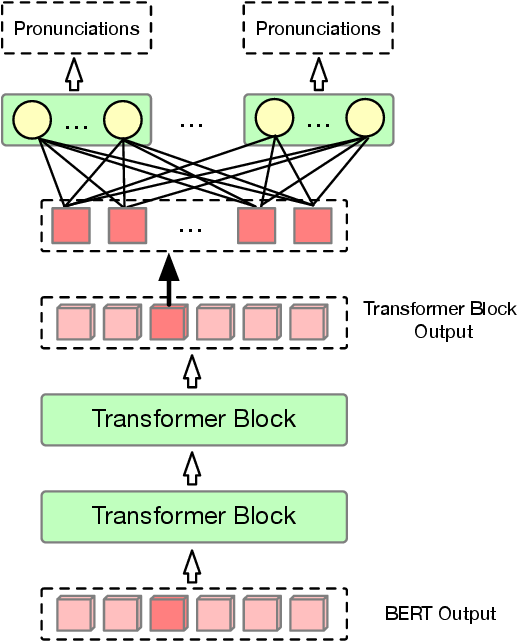}
		\centerline{(c)\quad\quad}\medskip
		
	\end{minipage}
\caption{The NN based classifiers, (a) fully-connected network based classifier, (b) LSTM based classifier, (c) Transformer block based classifier }
\label{fig:cls}
\end{figure*}

As the pre-trained BERT extracts semantic features from raw character sequence and the pronunciation of a polyphonic character is determined by its contextual semantics, we can directly predict a polyphonic character's pronunciation according to these semantic features. We assume that the polyphonic word is the $i$th element of the BERT input sequence, The NN based classifier predicts pronunciation based on BERT output and the value of the subscript $i$. We explored fully-connected network based classifier, LSTM based classifier and Transformer block based classifier in our research.

\subsubsection{Fully-connected network based classifer}

First of all, we use a two-layer fully-connected network to predict the pronunciation of polyphonic character according to the $i$th element of the BERT output sequence. The fully-connected network based classifier is depicted in Fig..\ref{fig:cls}-(a). The first fully-connected layer is shared by all the polyphonic characters. As for the second fully-connected (output) layer, it is not shared.  Each polyphonic character has a separate output layer whose units number is equal to the number of possible pronunciations. Softmax cross-entropy loss is adopted to train the classifier. The  LSTM based classifier and Transformer block based classifier also use the same structure of output layer and loss function.

\subsubsection{LSTM based classifer}

Indicated by \cite{shan2016bi}, contextual information such as the POS of polyphone's neighbor words can also affect the pronunciation of a polyphonic character.  So instead of classifying according to the $i$th element of Bert output sequence directly, we apply Bidirectional LSTM (BLSTM) to model the contextual information before classifying. The LSTM based classifier is shown in Fig.\ref{fig:cls}-(b). The BERT output sequence is processed by a two-layer BLSTM network first to model the contextual information, then a following unshared output layer predicts the pronunciation of corresponding polyphonic character according to the $i$th element of LSTM output sequence. 

\subsubsection{Transformer block based classifier}

Due to the characteristics of the recurrent network, nearby locations have a greater impact than farther locations. In order to better analyze the impact of context information, we use Transformer block to model context information. From the perspective of model structure, information at any position is equally important in Transformer block. The Transformer block based classifier depicted in Fig.\ref{fig:cls}-(c). Two-layer Transformer blocks model the contextual information on the BERT output, the following unshared output layer accepts the $i$th element of Transformer block as input and predicting the pronunciation of corresponding polyphonic character.

\section{Experiment and analysis}

\subsection{Dataset}

The experiments were conducted on a dataset extracted from TTS corpus in Tencent AI Lab. There  are 331,325 sentences containing polyphonic characters in the corpus. We selected polyphonic characters which appear in more than 2,000 sentences accounting for 83.7\% of the total polyphonic samples.  In our experiments, the dataset was divided into 10 subsets randomly keeping the distribution of polyphonic characters, 8 subsets were used for training, one subset was used as development set and the remaining subset as test set. We conducted 10-fold cross-validations to get the final average result.

\subsection{Settings of baseline approach}

We took Shan's LSTM approach for polyphone disambiguation  in \cite{shan2016bi} as baseline (LSTM baseline). The LSTM baseline approach needs word tokenization and POS tagging first on the input character sequence. Then the LSTM based model accepts a character embedding sequence and a contextual POS embedding sequence as input. The character embedding is generated from characters composing the word that contains polyphonic character. The POS sequence considering the neighbor words besides the word containing a polyphonic character.

\begin{figure}[htbp]
	\centering
	\includegraphics[scale=0.4]{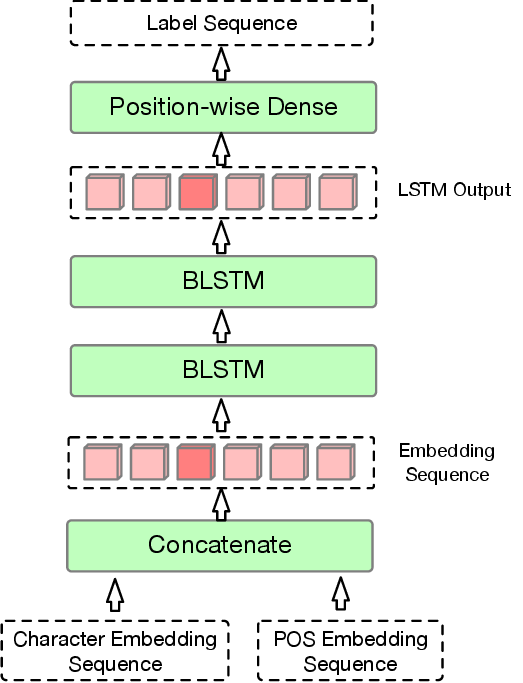}
	\caption{LSTM baseline approach for polyphone disambiguation}
	\label{fig:lstmbaseline}
\end{figure}

Fig.\ref{fig:lstmbaseline} depicts the LSTM baseline approach. The prediction of pronunciation is viewed as a sequence labeling task. The baseline model accepts embedding sequence concatenated by character embedding sequence and POS embedding sequence, and it outputs a label sequence corresponding to the input characters. In our experiments, we set the hidden units of BLSTM to 512, the number of BLSTM layers to 2 and the contextual size to 1 when constructing POS embedding sequence, identical to the setting in \cite{shan2016bi}.

\subsection{Settings of the proposed approach}

In our experiments, we adopted the pre-trained BERT model provided by Google to extract semantic features from raw Chinese character sequence\footnote{https://github.com/google-research/bert}, and the detail of BERT is identical to the $\rm BERT_{BASE}$ model described in \cite{devlin2018bert} whose output size of Transformer block is 768. Due to limited data, fine-tuning the pre-trained BERT model did not achieve desired results. So during the training phase, we froze the parameters of the BERT model and only updated the parameters of the NN based classifier.

In the fully-connected network based classifier, the hidden units of the first fully-connected layer is 512 with a dropout rate of 0.5 during training phase. The hidden units of BLSTM is  512 in LSTM based classifier. In Transformer block based classifier, The dimension of Transformer block is 512, and the head number is 8 in multi-head attention which is identical to settings in \cite{vaswani2017attention}. During the training phase, we adopted the Adam\cite{kingma2014adam} optimizer and set the learning rate to 5e-4, excepted the Transformer block based classifier. When training the classifier based on Transformer block, we use the training strategy described in \cite{vaswani2017attention}.

\subsection{Experimental results and analysis}

\begin{figure}[t]
	\centering
	\includegraphics[scale=0.4]{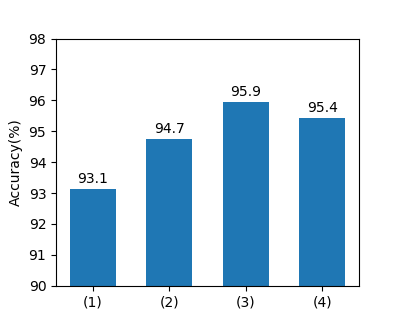}
	\caption{Experimental results: accuracies of defferent methods, (1) LSTM baseline, (2) BERT + FC, (3) BERT + LSTM, (4) BERT + Transformer block}
	\label{fig:acc}
\end{figure}

%\begin{figure}[htbp]
%	
%	\begin{minipage}[t]{0.2\linewidth}
%		\centering
%		\centerline{\includegraphics[scale=0.4]{pics/zhang1.png}}
%		\centerline{(a) \quad\quad}\medskip
%		
%	\end{minipage}%
%	\begin{minipage}[t]{0.2\linewidth}
%		\centering
%		\includegraphics[scale=0.4]{pics/di1.png}
%		\centerline{(b)\quad\quad}\medskip
%		
%	\end{minipage}
%
%	\caption{PCA embedding of BERT output corresponding to polyphonic character }
%	\label{fig:pca}
%\end{figure}

The experimantal result is depected in Fig.\ref{fig:acc}. All of our proposed methods outperforms the LSTM baseline, showing that the BERT model preprocessed on a large amount of unlabeled data effectively extracts semantic features, which greatly enhances the performance of pronunciation prediction. Comparing  \textit{ BERT + FC } with \textit{  BERT + LSTM} and \textit{ BERT + Transformer block}, We can draw the conclusion that the contextal information can really improve the performance for polyphone disambiguation. \textit{ BERT + LSTM} is better than \textit{ BERT + Transformer block}, indicating that contextual information in adjacent locations are more useful than in distant locations.\begin{CJK*}{UTF8}{gbsn}中文\end{CJK*}

\begin{figure}[t]
	\centering
	\includegraphics[scale=0.35]{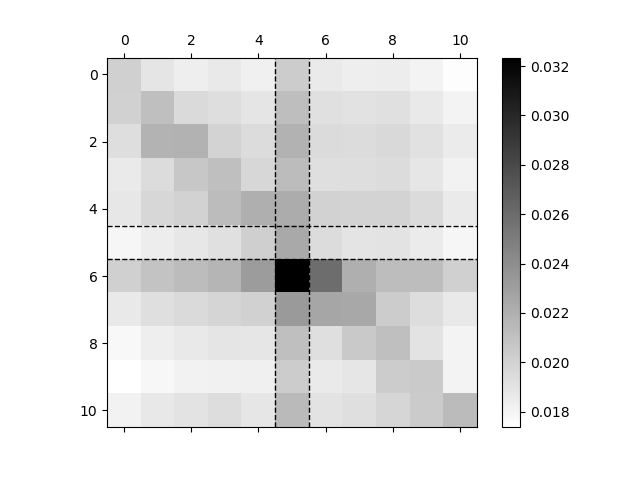}
	\caption{attention average weight cropped around polyphonic character}
	\label{fig:attenmap}
\end{figure}

To further illustrate the impact of contextual information for polyphone disambiguation, we draw the attention weights cropped around the polyphonic character. Fig.\ref{fig:attenmap} shows the average cropped attention weight of all the heads in the first Transformer block on test set. The attention weight is cropped around the polyphonic character with neighboring contextual size 5, which means the size of the cropped attention weight is (11, 11) and the location (5, 5) corresponding to the polyphonic character. It can be seen from Fig.\ref{fig:attenmap} that the closer to the position of the polyphonic character, the greater the weight of the attention, which indicates that the closer the context information is to the position of the polyphonic character, the more important it is for polyphone disambiguation. This also explains the reason why the LSTM based classifier is better than the Transformer block based one, LSTM can better model closer information due to the characteristics of the recurrent network.

%\subsection{Analysis}

\begin{figure}[t]
	\begin{minipage}[b]{.48\linewidth}
		\centering
		\centerline{\includegraphics[width=4.2cm]{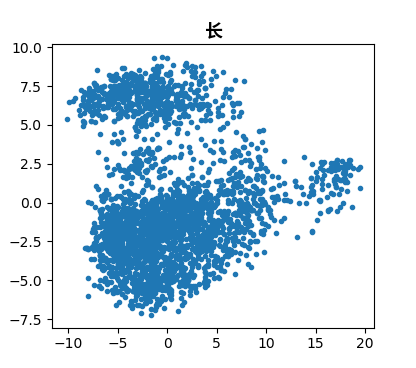}}
		%  \vspace{1.5cm}
		\centerline{(a)  }\medskip
	\end{minipage}
	\hfill
	\begin{minipage}[b]{0.48\linewidth}
		\centering
		\centerline{\includegraphics[width=4.2cm]{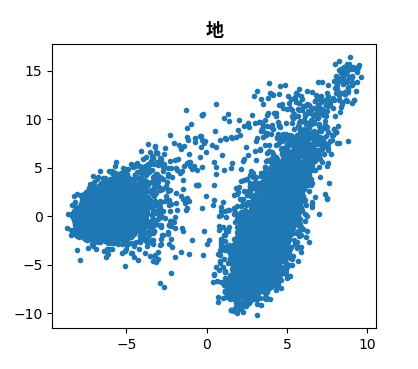}}
		%  \vspace{1.5cm}
		\centerline{(b)}\medskip
	\end{minipage}

	\caption{PCA embedding of semantic features extracted by BERT corresponding to Chinese character, (a)  ZHANG, (b) DI}
	\label{fig:pca}
\end{figure}

\section{Conclusions}

In this paper, we proposed an end-to-end framework for Chinese polyphone disambiguation. The proposed framework accepts raw Chinese character sequence as input without any preprocessing, and it consists of a pre-trained BERT model and a NN based classifier.

We implemented three classifiers based on neural network and conducted experiments together with the LSTM baseline \cite{shan2016bi}. The experimental results demonstrate that the BERT models pre-trained on a large amount of unsupervised data can effectively extract semantic features, which greatly enhances the performance of polyphone disambiguation. Meanwhile, the contextual information can also improve the result of polyphone disambiguation, especially the closer the context is to the polyphonic character, the greater its influence on polyphone disambiguation.

\bibliographystyle{IEEEtran}

\bibliography{mybib}

\end{document}